\documentclass{article}
\usepackage{spconf,amsmath,graphicx}

\usepackage{algorithm}
\usepackage{algorithmic}
\usepackage{amsmath}
\usepackage{graphicx}
\usepackage{multirow}
\usepackage{wrapfig}
\usepackage{color}
\usepackage{subcaption}
\usepackage{booktabs}
\usepackage{url}

\title{DenoiSpeech: Denoising Text to Speech with Frame-Level Noise Modeling}
\name{Chen Zhang$^{\star \dagger}$ \qquad Author Name$^{\star}$ \qquad Author Name$^{\dagger}$}
\name{Chen Zhang$^{1}$, Yi Ren$^{1}$, Xu Tan$^{2}$, Jinglin Liu$^{1}$, Kejun Zhang$^{1}$, Tao Qin$^{2}$, Sheng Zhao$^{3}$, Tie-Yan Liu$^{2}$}
\address{$^{1}$Zhejiang University, China, $^{2}$Microsoft Research Asia, $^{3}$Microsoft Azure Speech \\ 
\{zc99,rayeren\}@zju.edu.cn, xuta@microsoft.com, \{jinglinliu,zhangkejun\}@zju.edu.cn,\\ \{taoqin,sheng.zhao,tyliu\}@microsoft.com}

\begin{document}
\maketitle
\begin{abstract}
While neural-based text to speech (TTS) models can synthesize natural and intelligible voice, they usually require high-quality speech data, which is costly to collect. In many scenarios, only noisy speech of a target speaker is available, which presents challenges for TTS model training for this speaker.
Previous works usually address the challenge using two methods: 1) training the TTS model using the speech denoised with an enhancement model; 2) taking a single noise embedding as input when training with noisy speech. However, they usually cannot handle speech with real-world complicated noise such as those with high variations along time.
In this paper, we develop DenoiSpeech, a TTS system that can synthesize clean speech for a speaker with noisy speech data. In DenoiSpeech, we handle real-world noisy speech by modeling the fine-grained frame-level noise with a noise condition module, which is jointly trained with the TTS model. Experimental results on real-world data show that DenoiSpeech outperforms the previous two methods by 0.31 and 0.66 MOS respectively.
\end{abstract}
\section{Introduction}
\label{sec:intro}
Text to speech (TTS) aims to synthesize natural and intelligible voice from text, and has achieved great progress due to the advance of deep learning. Neural based TTS models can synthesize high-quality voice when training with a large amount of clean speech data. However, collecting clean speech data requires a quiet environment and good recording equipment such as those in professional speech studios, which incurs high collection cost. On the contrary, noisy speech is easy to collect, such as daily conversations and public talks. If we can leverage noisy speech data of a speaker for TTS model training in order to synthesize clean speech for this speaker, we will reduce a lot of data collection cost and greatly extend the applicability of TTS.

However, it is non-trivial to model the noisy speech data in TTS. A straight-forward way~\cite{valentini2016speech,valentini2016investigating,valentini2018speech,dai2020noise} is to first denoise the noisy speech with a pre-trained speech enhancement model~\cite{yin2020phasen}, and then use the enhanced speech to train the TTS model. Although this method usually performs well in simple noise situations, it performs much worse when the noise situation is very complicated and of different distributions from that used in training the enhancement model, and thus the \textit{enhanced} speech will in turn harm the TTS model training. Therefore, recent works~\cite{hsu2018hierarchical,hsu2019disentangling} proposed to directly train the TTS model using the noisy speech data. They provide the noise embedding of the corresponding noisy speech as conditional input for the noisy speech, and remove the noise information during inference to generate clean speech. However, the noise embedding they used is usually an utterance-level vector, which is coarse-grained and cannot cover the complicated noise patterns that vary greatly along the time dimension.

In this paper, we propose DenoiSpeech, which improves the previous methods with fine-grained frame-level noise modeling to better handle the complicated noise in real-world noisy speech. Specifically, we design a noise condition module which leverages a noise extractor to extract the frame-level noise information from the target-speaker noisy speech and then the extracted noise information is taken as input by the TTS decoder. The noise extractor is jointly trained with the TTS loss, as well as an adversarial CTC loss to ensure only noise information but not semantic information is extracted. We conduct experiments on both the artificial noisy dataset (mixed by VCTK corpus and NonSpeech100) and real-world noisy dataset (an internal noisy dataset) to evaluate our DenoiSpeech. Experimental results show that DenoiSpeech outperforms the previous works including those using utterance-level noise embedding and those leveraging a speech enhancement model, in both the artificial and real-world noisy datasets. Audio samples generated by DenoiSpeech can be found at \url{https://speechresearch.github.io/denoispeech}.

\section{Method}

\begin{figure*}[!thb]
	\centering
	\begin{subfigure}[h]{0.25\textwidth}
	    \captionsetup{justification=centering}
		\centering
		\includegraphics[width=\textwidth,trim={0.1cm 4.32cm 21.7cm 4cm}, clip=true]{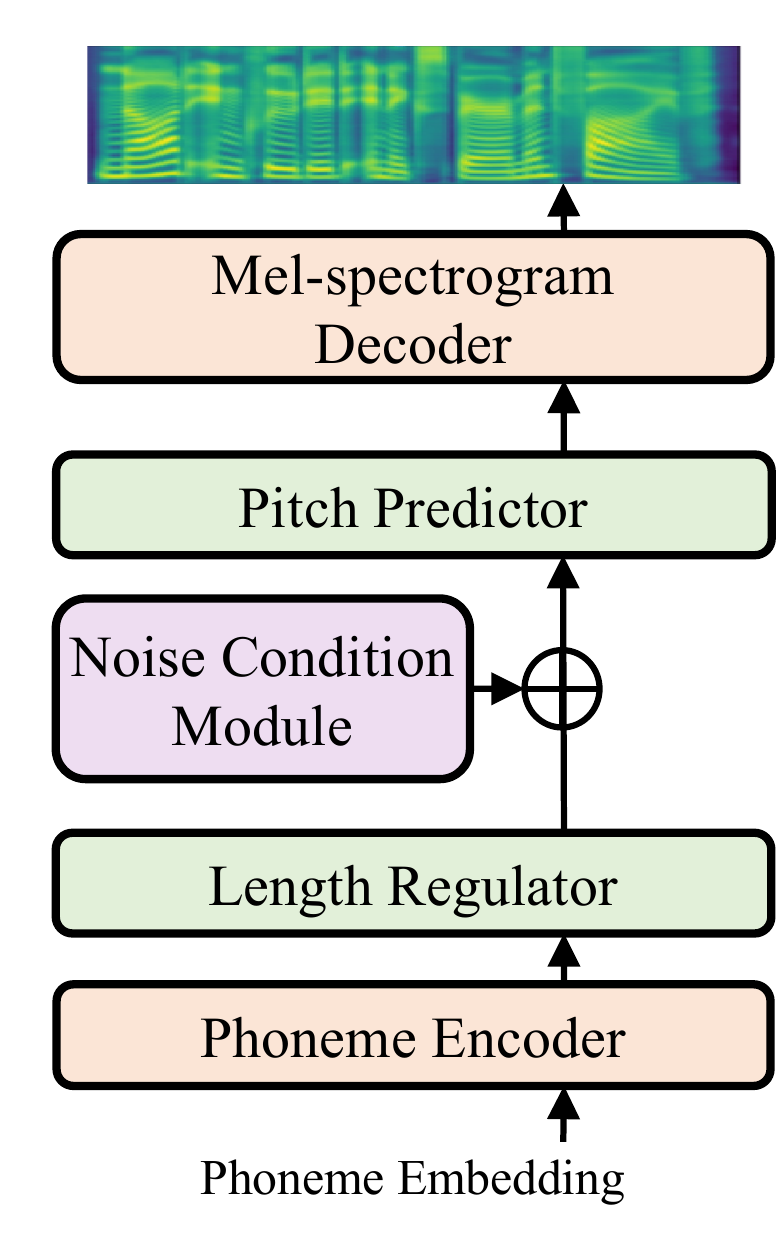}
		\vspace{-0.4cm}
		\caption{DenoiSpeech}
		\label{fig:pipeline}
	\end{subfigure}
	\hspace{0.15cm}
	\begin{subfigure}[h]{0.21\textwidth}
	    \captionsetup{justification=centering}
		\centering
		\includegraphics[width=\textwidth,trim={8.2cm 4.32cm 14.8cm 4cm}, clip=true]{figs/new_archs.pdf}
		\vspace{-0.3cm}
		\caption{Noise Condition Module}
		\label{fig:arch_noise_condition_module}
	\end{subfigure}
	\hspace{0.15cm}
	\begin{subfigure}[h]{0.21\textwidth}
	    \captionsetup{justification=centering}
		\centering
		\includegraphics[width=\textwidth,trim={14.9cm 4.32cm 8.1cm 4cm}, clip=true]{figs/new_archs.pdf}
		\vspace{-0.31cm}
		\caption{Noise Extractor}
		\label{fig:arch_noise_extractor}
	\end{subfigure}
	\hspace{0.15cm}
	\begin{subfigure}[h]{0.24\textwidth}
	    \captionsetup{justification=centering}
		\centering
		\includegraphics[width=\textwidth,trim={21.7cm 4.32cm 0.1cm 4cm}, clip=true]{figs/new_archs.pdf}
		\vspace{-0.1cm}
		\caption{Noise Encoder}
		\label{fig:arch_noise_encoder}
	\end{subfigure}
	\caption{The overall architecture for DenoiSpeech.}
	\label{fig:noise_extractor}
	\vspace{-0.5cm}
\end{figure*}
\label{method}
In this section, we first give an overview of the model architecture in DenoiSpeech and then describe our proposed noise condition module. Finally, we introduce the training and inference procedure of our DenoiSpeech.

\subsection{Model Overview}
The overall model architecture of DenoiSpeech is shown in Figure~\ref{fig:pipeline}, which is based on FastSpeech 2~\cite{ren2020fastspeech}. The phoneme encoder converts the phoneme embedding into the hidden sequence, and then the length regulator extends the sequence to the same length as the mel-spectrogram sequence. Then, the noise condition module extracts the noise condition from the noisy speech and adds it into the hidden sequence. We then use pitch predictor to add pitch into the hidden sequence as in FastSpeech 2. Finally, the mel-spectrogram decoder converts the hidden sequence into mel-spectrogram sequence in parallel. We describe the designs of the noise condition module in the next subsection.

\subsection{Noise Condition Module}
The noise condition module aims to capture the noise information in noise speech, which is then used as input of the TTS model. As shown in Figure~\ref{fig:arch_noise_condition_module}, the noise extractor extracts the background noise audio, which is then converted into noise condition using the noise encoder. After that, the noise condition is added to the hidden sequence to introduce the noise information to our model~\footnote{We observe that background noise can affect  the extracted pitch of the speech, and we add the noise condition module before the pitch predictor so that the pitch predictor can take the noise information into consideration, which is more reasonable and proved to be more effective by the experimental results.}. The adversarial CTC sub-module in Figure~\ref{fig:arch_noise_condition_module} is introduced to ensure only noise information is extracted, without any text/phoneme related information.

\subsubsection{Noise Extractor}
The noise extractor aims to extract the noise audio $y'$ from noisy speech $y$.
As shown in Figure~\ref{fig:arch_noise_extractor} , our noise extractor is based on UNet~\cite{ronneberger2015u}, which is widely used in speech enhancement~\cite{pascual2017segan}. There are 4 DownConv blocks and 4 UpConv blocks in the noise extractor. Each DownConv/UpConv block consists of a downsampling/upsampling layer and the repeated application of two 3x3 2D-convolutional layers, each followed by a ReLU activation function and a batch normalization layer. Our noise extractor is trained with two kinds of data, which are described in Section~\ref{sec:datasets}: 1) paired noisy data $(y_p, y’_p)$, where $y’_p$ is the noise sequence and $y_p$ is the noisy speech sequence (which is usually constructed artificially by mixing clean speech and noise $y’_p$); 2) unpaired noisy data $y_u$, where the noisy speech has no paired noise (which can be artificial or real-world noisy data). To leverage the unpaired noisy data into training, as shown in Algorithm~\ref{alg:denoise_tts}, we jointly train the noise extractor with the TTS model, which can help provide end-to-end gradient to optimize the noise extraction.

\subsubsection{Noise Encoder}
Our noise encoder converts the noise audio into fine-grained frame-level noise condition. As shown in Figure~\ref{fig:arch_noise_encoder}, the noise audio is encoded into the noise condition, and then the noise condition is added to the hidden sequence which is then fed into the pitch predictor. The fine-grained frame-level noise condition has the same length as the output mel-spectrogram and thus can describe the noise information of each timestep, which helps our DenoiSpeech handle the noise with larger variance along time. We use the ground truth noise audio $y’_p$ as the input of the noise encoder for the paired noisy speech, and the extracted noise audio by the noise extractor for the unpaired noisy speech. We also use clean speech for TTS model training; for clean speech, we use silence audio (with the same length as the clean speech) as the input of the noise encoder.

\subsubsection{Adversarial CTC Module}
If the noise generated from the noise extractor contains text information, there is information leakage and thus would make the TTS model training trivial. To ensure the noise extractor can generate only noise information instead of text information, we add an adversarial CTC module after the noise extractor as shown in Figure~\ref{fig:arch_noise_condition_module}. The noise audios extracted from the unpaired noisy speech are fed into the Gradient Reverse Layer (GRL) and the CTC encoder to calculate the adversarial CTC loss. 
In this way, we force the extracted noise audios to not contain speech content information. The CTC encoder is based on a Transformer encoder with an additional linear-softmax layer to project the hidden states to character-level output distribution.

\subsection{Training and Inference}
Finally, we describe the training and inference procedure of DenoiSpeech according to the description in the previous subsections. The detailed procedure is shown in Algorithm~\ref{alg:denoise_tts}.

\vspace{-0.3cm}
\begin{algorithm}[htb]
\small
\caption{Training and Inference of DenoiSpeech}
\label{alg:denoise_tts}
\begin{algorithmic}
   \STATE {\bfseries Training:}
   \STATE {\bfseries Input:} Paired noisy speech corpus $(\mathcal{Y}_p, \mathcal{Y}_p')$, unpaired noisy speech corpus $\mathcal{Y}_u$, clean speech corpus $\mathcal{Y}_c$. Paired phoneme and speech corpus $(\mathcal{X}, \mathcal{Y})$ where $\mathcal{Y}$ consists of $\mathcal{Y}_p$, $\mathcal{Y}_u$ and $\mathcal{Y}_c$. 
   \STATE {\bfseries Step 1:} Train the noise extractor using corpus $(\mathcal{Y}_p, \mathcal{Y}_p')$ for certain steps. 
   \STATE {\bfseries Step 2:} Jointly train the TTS model and noise condition module on $(\mathcal{X}, \mathcal{Y})$ with the warmstarted noise extractor. 
\\\hrulefill 
   \STATE {\bfseries Inference:}
   \STATE {\bfseries Input:} A phoneme sequence, silence audio, the trained DenoiSpeech model.
   \STATE {\bfseries Step:} Take the silence audio as the input of noise encoder to synthesize clean speech of the phoneme sequence. 
\end{algorithmic}
\end{algorithm}
\vspace{-0.5cm}

\section{Experiments and Results}
\label{exp}
In this section, we first introduce the experimental setup, and then report the results of DenoiSpeech. Finally, we conduct some analyses on DenoiSpeech.

\subsection{Experimental Setup}
\label{sec:datasets}
\textbf{Datasets.} We use two types of datasets: artificially generated noisy dataset and real-world noisy dataset.
For artificial noisy dataset, we use the VCTK corpus~\cite{veaux2016superseded} as the clean speech and the Nonspeech100~\cite{hu2010tandem} for background noise. The VCTK corpus contains 44 hours clean English speech and corresponding text with 109 speakers, and Nonspeech100 contains 100 noise utterances from 20 different noise classes. 
Following~\cite{hsu2019disentangling}, half of the speakers in VCTK are randomly selected to be noisy to simulate speaker-correlated noise.
All of the utterances of the noisy speakers are mixed with noise sampled from Nonspeech100 with an SNR randomly chosen from 5 - 25 dB. 
We split the noisy speech into two parts: paired noisy data and unpaired noisy data, which have totally different speakers and background noise.
For the real-world noisy dataset, we use an internal noisy dataset collected from real-world noisy situations, which contains about 25 hours noisy English speech from 55 speakers. This real-world noisy dataset contains more complicated noise types, not only including the additive noise we simulated, but also reverberation and multiplicative noise, whose patterns and distributions are quite different from those in the artificial dataset. 
We first convert the text in the VCTK corpus and the real-world noisy dataset into phoneme with an open-sourced grapheme-to-phoneme conversion tool\footnote{https://github.com/bootphon/phonemizer}, and then we extract the phoneme duration with MFA~\cite{mcauliffe2017montreal}. We convert all the speech waveform with the sample rate of 22050 in our experiments into mel-spectrogram
following~\cite{ren2020fastspeech} with a frame size of 50 ms and hop size of 12.5ms.

\textbf{Model Configuration.} We choose FastSpeech 2~\cite{ren2020fastspeech} as the basic model structure since it can synthesize high-quality speech with high speed. And our method can also be applied to other end-to-end TTS models. 
We stack $N=4$ layers of Transformer blocks in both the text encoder and mel-spectrogram decoder. In each Transformer block, the hidden size of both the self-attention and feed-forward layers and the dimension of the phoneme embeddings are set to $256$, and the filter size of the feed-forward layer is set to $1024$. For the noise extractor, we choose UNet~\cite{ronneberger2015u} as the basic model structure following ~\cite{pascual2017segan}, since this structure shows good performance on both speech enhancement task and source separation task, which are similar to our task.


\textbf{Training Setup.}
First, we train the noise extractor for 40K steps with other model parameters fixed on 1 GeForce RTX 2080 Ti GPU, and set the batch size to 12 sentences. After that, we train the whole DenoiSpeech for 120K steps.
Following ~\cite{ren2020fastspeech}, we use Adam optimizer ~\cite{kingma2014adam} and set $\beta_1$, $\beta_2$, $\varepsilon$ to $0.9$, $0.98$ and $10^{-9}$ respectively. We use the sum of mean absolute error (MAE) and mean structural similarity (MSSIM)~\cite{wang2004image} losses for mel-spectrogram value prediction inspired by ~\cite{vainer2020speedyspeech}. We train the adversarial CTC module with connectionist temporal classification (CTC) loss~\cite{graves2006connectionist}. For the length regulator, we use mean square error (MSE) loss and apply MAE loss for both the noise extractor and the pitch predictor. 

\textbf{Inference and Evaluation.}
\label{infer_and_eval}
In inference, we use the silence audio as the input of the noise encoder in the noise condition module to generate the mel-spectrogram of noisy speakers without background noise. For evaluation, we use the mean opinion score (MOS)~\cite{chu2006objective} to measure the perceptual quality of synthesized speech: each synthesized speech sample is judged by 20 native speakers. 

\subsection{Results}
\label{sec:main_results}
We evaluate the quality of output audio of DenoiSpeech. We compare the MOS of the audio samples generated by DenoiSpeech with other systems: 1) \textit{Clean GT}, the ground-truth clean recordings; 2) \textit{Noisy GT}, the ground-truth clean recordings mixed with background noise in the artificial noisy corpus or the ground-truth noisy recordings of real-world noisy corpus; 3) \textit{Augment-Adversarial}, apply data augmentation and adversarial factorization to FastSpeech 2 as used in~\cite{hsu2019disentangling} ; 4) \textit{Enhancement-Based}, FastSpeech 2 trained on enhanced speech denoised with an enhancement model~\cite{yin2020phasen}. Systems in 3) and 4) use Parallel WaveGAN~\cite{yamamoto2020parallel} as the vocoder. As shown in Table~\ref{tab:main_results}, we conduct experiments both on the artificial noisy corpus and the real-world noisy corpus. DenoiSpeech greatly outperforms \text{Noisy GT} and generate more natural and clearer speech than two baselines, which demonstrates the effectiveness of fine-grained frame-level noise condition in DenoiSpeech. It can be seen that the gaps between two baselines and DenoiSpeech on the real-world noisy corpus are bigger than those on the artificial noisy corpus, which verifies the robustness and effectiveness of DenoiSpeech on real-world noisy corpus.

\begin{table}[!tbh]
\centering
\small 
\vspace{-0.3cm}
\begin{tabular}{ l | c  c }
	\toprule
	Method &  Artificial  & Real-World \\
	\midrule
	\textit{Clean GT} & 4.16 & --\\
    \textit{Noisy GT} & 2.69 & 3.02\\
	\midrule
	\textit{Augment-Adversarial~\cite{hsu2019disentangling}} & 3.43 & 2.65\\
	\textit{Enhancement-Based~\cite{yin2020phasen}} & 3.66 & 3.04\\
	\midrule
	\textit{DenoiSpeech} & \textbf{3.77} & \textbf{3.35}\\
	\bottomrule
\end{tabular}
\vspace{-0.2cm}
\caption{The MOS of DenoiSpeech and previous systems.}
\label{tab:main_results}
\vspace{-0.5cm}
\end{table}

\subsection{Method Analyses}
We conduct more experimental analyses on our methods in this subsection
to verify the necessity of some design details. 
\subsubsection{The Granularity of Noise Condition}
We study how the granularity of the noise condition can influence the synthesized speech quality. We take the average of the frame-level noise condition along the time dimension as the utterance-level noise condition. And we also conduct experiments without noise condition. As shown in Table~\ref{tab:granularity}, frame-level noise condition outperforms utterance-level one by -0.59 CMOS, which demonstrates that fine-grained noise conditions can describe the noise information better.

\begin{table}[!tbh]
\vspace{-0.3cm}
    \centering
    \begin{tabular}{c|c}
        \toprule
		Setting &  CMOS \\
		\midrule
        \textit{Frame-Level Noise Condition} & 0 \\
        \textit{Utterance-Level Noise Condition} & -0.59 \\
        \textit{Without Noise Condition} & -0.90 \\
	    \bottomrule
    \end{tabular}
    \vspace{-0.2cm}
    \caption{The CMOS of different noise condition granularity.}
    \label{tab:granularity}
    \vspace{-0.5cm}
\end{table}

\subsubsection{Joint Training for Noise Extractor}
\label{sec:fixed_noise_extractor}
We study how joint training improves performance. We compare DenoiSpeech with another model that fixes the noise extractor in Step 2 in Algorithm~\ref{alg:denoise_tts}. As shown in Table~\ref{tab:two_stages}, the quality of the synthesized speech drops when we fix the noise extractor during the TTS model training, which shows that jointly training the noise extractor and the TTS model can have better generalization performance.

\begin{table}[!tbh]
\vspace{-0.3cm}
    \centering
    \begin{tabular}{c|c}
        \toprule
		Setting &  CMOS \\
		\midrule
        \textit{Jointly Noise Extractor} & 0 \\
        \textit{Fixed Noise Extractor} & -0.16 \\
	    \bottomrule
    \end{tabular}
    \vspace{-0.2cm}
    \caption{The CMOS of fixed noise extractor.}
    \label{tab:two_stages}
    \vspace{-0.8cm}
\end{table}
\subsubsection{Adversarial CTC Module}
In order to study how the adversarial CTC module influences the model performance, we remove the adversarial CTC module and the comparison results are listed in Table~\ref{tab:adv_ctc}. As the results show, the adversarial CTC module improves audio quality, indicating the effectiveness of the adversarial CTC module in extracting the noise audio.
\begin{table}[!tbh]
    \centering
    \vspace{-0.3cm}
    \begin{tabular}{c|c}
        \toprule
		Setting &  CMOS \\
		\midrule
        \textit{With Adversarial CTC Module} & 0 \\
        \textit{Without Adversarial CTC Module} & -0.09 \\
	    \bottomrule
    \end{tabular}
    \vspace{-0.2cm}
    \caption{The CMOS comparison of adversarial CTC module.}
    \label{tab:adv_ctc}
    \vspace{-0.8cm}
\end{table}


\section{Conclusion}
In this work, we proposed DenoiSpeech, a denoising text-to-speech system with frame-level noise condition, to synthesize clean speech for the speaker with only noisy training data. DenoiSpeech introduces a noise condition module to provide the corresponding noise information for noisy speech training, where the noise condition module consists of a noise extractor to extract noise information from noisy speech and an adversarial CTC module to ensure the noise extractor not to generate text information so as to avoid information leakage.
Our experimental results show that DenoiSpeech outperforms other methods on both the artificial noisy corpus and real-world noisy corpus in terms of synthesis quality. Our further method analyses verify the effectiveness of each module in DenoiSpeech. In the future, we will consider training TTS model on more diverse types of noise and apply few-shot learning strategies on noisy speakers.

\bibliographystyle{IEEEbib}
\bibliography{refs.bib}

\end{document}